\documentclass[aps,prapplied,reprint,superscriptaddress]{revtex4-2}
\usepackage[utf8]{inputenc}
\usepackage[english]{babel}
\usepackage[T1]{fontenc}
\usepackage{amsmath}
\usepackage{hyperref}

\usepackage{graphicx}
\usepackage{dcolumn}
\usepackage{bm}
\usepackage{comment}
\newcommand{\myvar}{\mathrm{var}}
\newcommand{\mycov}{\mathrm{cov}}

\usepackage{physics}
\usepackage{amsmath}
\usepackage{float}
\usepackage[section]{placeins}

\usepackage{xcolor}

\newcommand{\new}[1]{{\textcolor{black}{ #1}}}
\newcommand{\ver}[1]{{\color{black}{ #1}}}

\newcommand{\rev}[1]{{\color{black}{ #1}}}

\newcommand{\re}[1]{{\color{black}{ #1}}}
\begin{document}

\title{Non-Gaussian state teleportation with a nonlinear feedforward}

\author{Vojt\v{e}ch Kala}
  \email{kala@optics.upol.cz}
  \affiliation{
 Department of Optics, Palack\'y University, 17. listopadu 1192/12, 77146 Olomouc, Czech Republic}%
\author{Mattia Walschaers}
\email{mattia.walschaers@lkb.upmc.fr }
\affiliation{
 Laboratoire Kastler Brossel, Sorbonne Universit\'e, CNRS, ENS-PSL Research University,
Coll\`ege de France; 4 place Jussieu, F-75252 Paris, France
}%
\author{Radim Filip}
\email{filip@optics.upol.cz}
\affiliation{
 Department of Optics, Palack\'y University, 17. listopadu 1192/12, 77146 Olomouc, Czech Republic}%
\author{Petr Marek}
\email{marek@optics.upol.cz}
\affiliation{
 Department of Optics, Palack\'y University, 17. listopadu 1192/12, 77146 Olomouc, Czech Republic}%

\begin{abstract}
Measurement-induced quantum computation with continuous-variable cluster states utilizes  \rev{teleportation to transmit and alter quantum states via measurement-and-feedforward control. One of the key challenges of this approach is the deterioration of quantum states caused by the noise added due to imperfect entanglement of the cluster. We analyze the propagation of a quantum non-Gaussian state with nonlinear squeezing through a small cluster state. We show that a nonlinear feedforward in the deterministic teleportation protocol reduces the added noise and improves the nonlinear squeezing transferred.} In a probabilistic regime, the improvement can be manifested even with current experimental resources. Better processing of non-Gaussian states can bring us closer to the necessary interplay between cluster states and non-Gaussianity required by quantum computing.

\end{abstract}

\maketitle


\section{Introduction}
\rev{The capacity of Gaussian cluster states for massive multi-mode multiplexing  \cite{OBrien2007,Yokohama2013,Larsen,Menicucci} is the main advantage of measurement-induced quantum computing with bosonic modes of traveling light. Accompanied by non-Gaussian measurements, their simulation complexity exceeds the abilities of classical computation \cite{GBS,Chabaud2023}. The main underlying working principle of the processing with the Gaussian cluster states is quantum teleportation. Measurements of the individual modes \rev{enable manipulating the} quantum information as it propagates along the cluster \cite{ClusterCV,Menicucci,Walshe2020}. The choice of measurement determines whether the state is simply transmitted by quantum teleportation \cite{Ban2002,Furusawa1998,Pirandola2015,Walshe2020,Takeoka} or transformed in a required manner \cite{Sefi2019,Filip2005,Miyata2016,Hillmann2022}. To reach the full advantage of the quantum computation, it is necessary to employ deterministic non-Gaussian operations \cite{Lloyd1999,Gottesman2001,Mari2012,Chabaud2023,Ohlinger} along with the Gaussian entangled cluster state. Experimental challenges motivated measurement-induced implementation of those operations, \cite{Miyata2016,Konno2021b} leaving two distinct tasks: experimental generation of the non-Gaussian states and their successful propagation through a Gaussian circuit.}

In quantum optics with traveling light, non-Gaussian quantum states have been generated in a wide range of experimental configurations \cite{Yukawa2013b,URen2004,Ogawa2016,Zavatta2004,Ourjoumtsev2006,LachmanG19}. Today, photon-number-resolving detectors, together with Gaussian operations such as parametric down-conversion and coherent classical displacements, are used to probabilistically generate non-Gaussian states with varied nonclassical and non-Gaussian properties \cite{Konno2021,Takeoka2011,Asavanant17,Melalkia22,Tiedau,Konno2024}. 
The basic principle of the state preparation, closely aligned to the Gaussian boson sampling problem \cite{GBS}, is still actively investigated to prepare \rev{more complex} quantum states \cite{Takase2023}. 

\rev{The successful generation of non-Gaussian states led to their active use in tests of their propagation through Gaussian circuits, resulting in the demonstration of key protocols. One of the most fundamental is quantum teleportation \cite{Ide,Mista,Fuwa,Takeda,Benichi,Lee,Zhao2023}, which is the cornerstone of measurement-based quantum computing \cite{Menicucci}.  The key challenge it is facing is the deterioration of quantum states due to the noise caused by imperfect quantum resources and as well experimental imperfections. This deterioration can be alleviated by adaptation - either of the protocol \cite{Eaton2022}, or of the quantum states \cite{Jeannic2018,Kala22,Provaznik2025}. The crucial decision that needs to be made is which of the features of a quantum state should be protected, as it is too challenging to protect all. Fidelity and Wigner function negativity enable to show average quality and necessary non-Gaussianity of a state processed by the quantum protocol, however, they are not sufficiently operational. Here we use the more practical notion of nonlinear squeezing. It appears as a perfect choice due to its operational definition, clear and straightforward interpretation in continuous variable quantum information protocols, and relevance for current experiments. As such, it enables us to proceed towards the concepts defined with ideal resources \cite{Konno2021b,Baragiola2019}. }

\rev{The nonlinear squeezing is defined as a noise suppression in a nonlinear function of quadratures that falls below a Gaussian threshold. The nonlinear squeezing is a resource for the deterministic realization of cubic phase gate \cite{Miyata2016}, which appears in the continuous variable quantum information in two contexts, as a part of the non-Clifford operation on GKP qubits \cite{Gottesman2001} and as an operation unlocking a universal control of a continuous mode \cite{Lloyd1999}. When a measurement-induced cubic operation is implemented via cubic nonlinear phase measurement \cite{sakaguchi}, the nonlinear squeezing is a single property required from an \rev{auxiliary} state. It directly determines the amount of unwanted noise added to the target result. Simultaneously, it is a specific example of a purely continuous-variable sign of non-Gaussianity \cite{Miyata2016,Kala22,Brauer2021}, relevant to current experimental techniques \cite{Konno2021} and different to e.g. Fock or Schr\"{o}dinger cat states \cite{Ourjoumtsev2006,Tiedau}. 

In this paper, we study \rev{the measurement-induced} propagation of the cubic nonlinear squeezing along a single node of a cluster state.  We consider the nonlinear squeezing as both the resource of non-Gaussianity and the figure of merit characterizing the processed state. We evaluate the performance of the canonical teleportation protocol optimized for nonlinear squeezing transfer and show its limits. Importantly, we also show that it is possible to go beyond these limits when the teleportation is enhanced by using the nonlinear feedforward \cite{Miyata2016,sakaguchi}. Finally, we show the ultimate limits on teleportation of cubic squeezing provided by remote state preparation by ideal cubic measurement. All these results provide a vital insight, necessary for further extensions of such measurement-induced protocols with large cluster states.}

\section{Teleportation, linear and nonlinear}

\subsection{Nonlinear squeezing}

\rev{The original role of the non-Gaussian \rev{auxiliary} state with the cubic nonlinear squeezing is to suppress a noise term that appears at the outcome of the measurement-induced cubic phase gate \cite{Miyata2016,Konno2021}
\begin{equation}\label{O}
   O =  p + z x^2,
\end{equation}}
the real parameter $z$ is called cubicity.
Inspired by \rev{the} insufficiency of Gaussian states for the quantum computing advantage with CV, the nonlinear squeezing is defined as a ratio of \rev{the} variance of the operator $O$ compared to the minimal variance that can be achieved in a Gaussian state \cite{Kala22,Brauer2021,Konno2021}
\begin{equation}\label{nlsq}
        \xi(z)_{\rho}=\frac{\myvar_\rho(p+zx^2)}{\min_G \myvar(p+zx^2)}.
\end{equation}
As a consequence of this definition, a state with nonlinear squeezing  \rev{($\xi(z)<1$)} is certainly non-Gaussian. \rev{The nonlinear squeezing can also be expressed in dB as 
\begin{equation}\label{xiindB}
    \xi(z)[\textrm{dB}] = 10 \log_{10}(\xi(z)).
\end{equation}}
\rev{Our goal, from now on, will be to analyze the quality of the nonlinear squeezing transferred through the cluster state. Simultaneously, the nonlinear squeezing injected into the cluster state serves as the only source of non-Gaussianity here. Its evaluation at the output is a figure of merit of the procedure quality.}

\rev{The maximal nonlinear squeezing is reached by the unphysical infinite energy cubic state
\begin{equation}\label{idfinf}
    \ket{\chi_{id}}=C(\chi)\ket{p=0},
\end{equation}
\rev{for $\chi = z$.
The eigenstate $\ket{p=0}$  of the $p$ quadrature is transformed by the cubic nonlinearity
\begin{equation}\label{kubik}
    C(\chi)=\exp(-i\chi\frac{x^3}{3}).
\end{equation}}
Such a state is well approximated by a cubically transformed squeezed vacuum
\begin{equation}\label{idf}
    \ket{\chi}=C(\chi)S(r)\ket{0},
\end{equation}
where the squeezed vacuum approximates the quadrature eigenstate.} This finitely nonlinearly squeezed state has finite energy and minimizes uncertainty relations. However, it is an idealized approximation in \rev{the} sense that its preparation still includes experimentally unfeasible cubic nonlinearity.

Experimentally feasible approximations of the finitely nonlinearly squeezed states can be prepared on a finite Fock subspace \cite{Yukawa2013b,Huang}. The approximations can be written as
\begin{equation}\label{superpo}
    \ket{\phi_N} = \sum_{n=0}^N c_n \ket{n},
\end{equation}
where the coefficients can be found \rev{efficiently} by solving an eigenvalue problem and two variable optimization \cite{Miyata2016}. Note that Fock states and states with nonlinear squeezing have different symmetries. Unlike the nonlinearly squeezed states that are symmetric in $x$ only, \rev{the Wigner functions of Fock states are symmetric in $x$ and $p$}. This difference hints that an approximation of the cubic state must be a superposition of Fock states. 
The experimentally feasible quantum state exhibiting nonlinear squeezing and simultaneously the lowest order approximation is a superposition of the first two Fock states
\begin{equation}\label{approx2}
    \ket{\phi_{1}}=u\ket{0}+i\sqrt{1-u^2}\ket{1},
\end{equation}
where $u$ is real and bounded by $0<u<1$ \cite{Konno2021}. The used parameterization ensures symmetry under transformation $x\rightarrow -x$ which holds also for \eqref{O}. One step further leads to \rev{auxiliary} state approximation
\begin{equation}\label{approx3}
    \ket{\phi_{2}}=c_0\ket{0}+ic_1\ket{1}+c_2\ket{2},
\end{equation}
where the $c_i$ are real coefficients. The states can be found as eigenstates of the operator $O$ restricted to an $N$ dimensional subspace of the Fock space. They provide the best nonlinear squeezing available on such a subspace \cite{Miyata2016}. \rev{In the further analysis, we employ those states as the resource for non-Gaussian measurements and operations.}
When processing a quantum state or seeking nonlinear squeezing after a nonlinear measurement on an entangled mode, the nonlinear squeezing of the output state does not need to be minimal for the initial cubicity. Thus we define the native cubicity $z_n$ of a given state $\rho$ as the cubicity for which 
\begin{equation}\label{nativez}
    \xi(z_n)_{\rho} =  \min_z \xi(z)_{\rho}.
\end{equation}
\ver{The native cubicity $z_n$, for which the state reveals its maximal nonlinear squeezing, can be considered a property of the state.}

\subsection{Deterministic regime}
The teleportation protocol \cite{Furusawa1998,Pirandola2015} is the basic building block of cluster state computing. \rev{Usually, a homodyne measurement is used together with a linear feedforward. The outcoming state is processed via a Gaussian displacement operation that linearly depends on the homodyne measurement results.} \rev{Following the idea of cluster state computing, it is possible to consider also generalized measurements}, that enable not only \rev{the} transfer of quantum states but also their transformation \rev{\cite{Filip2005,Menicucci}}. Such measurements can include \rev{an} adjustable nonlinear feedforward that utilizes the measurement results in some more complex function. \re{In the following, we discuss the nonlinear cubic transformation as the ideal fundamental limit, its implementation using the nonlinear feedforward as an experimentally available transformation and the canonical teleportation as the standard protocol for comparison. }

\begin{figure*}[htbp]
\centering
\includegraphics[width=0.9\textwidth]{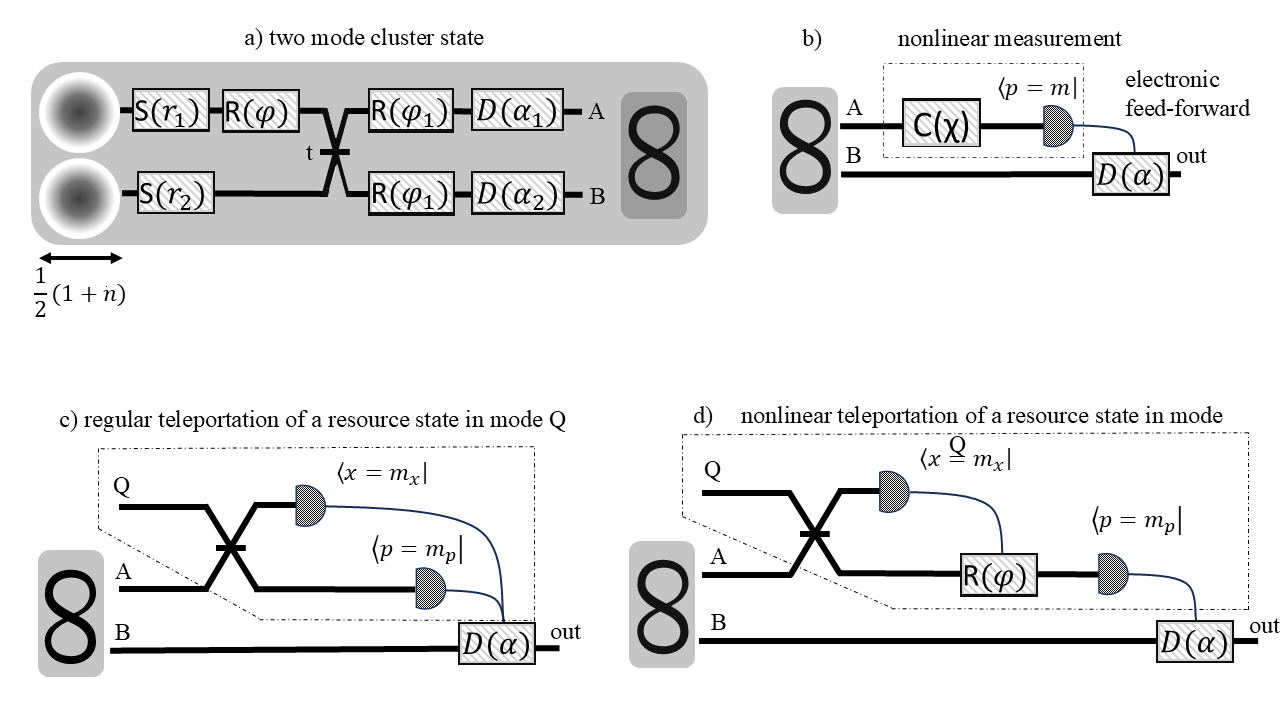}
\caption{a) \rev{Optical scheme illustrating the considered cluster state generation.} b) Optical scheme showing the nonlinear cubic phase measurement applied to one part of the cluster state \rev{to project it onto the cubic state}. The measurement is decomposed into an ideal cubic nonlinearity followed by homodyne measurement. The measurement result is used in a feedforwarded displacement on the second mode. c) Regular teleportation scheme \rev{with linear feedforward} transferring the quantum state through the \rev{optical} cluster state. d) Nonlinear teleportation with nonlinear feedforward and \rev{auxiliary} state in the mode Q. }
\label{Schemata}
\end{figure*}

A single step of the transfer, the canonical teleportation protocol itself, is schematically depicted in Fig. \ref{Schemata}c. The initial state in mode $Q$ interacts on a balanced beam splitter with mode $A$ of a two-mode entangled state. \rev{The modes $Q$ and $A$} are measured by a pair of homodyne detectors. The measured values are then used for the feedforward displacement operation on the remaining mode $B$ of the entangled state. \rev{Altogether, the canonical teleportation protocol transforms the initial quadrature operators as}
\begin{equation}\label{Htelep}
    \begin{split}
        x_B^{\textrm{out}} =& x_Q + x_B-x_A=x_Q+N_x\\
        p_B^{\textrm{out}} =& p_Q+p_A+p_B=p_Q+N_p.
    \end{split}
\end{equation}
The expressions for individual transformations generated by beam splitter and other optical elements can be found in Appendix \ref{Heisenbergtrasformations}. Quadratures $x$ and $p$ of the optical field obey $[x,p]=i$. The statistics of the teleported state \rev{are} accompanied by noise terms $N_x$ and $N_p$, \rev{which depend} on the properties of the two-mode entangled state, illustrated in 
Fig.~\ref{Schemata}a, and vanish for the ideal EPR state with maximal correlation and infinite energy. Despite the protocol yielding perfect results in the limit of infinite squeezing, the real performance will always be noisy, with an imperfect outcome.

\rev{Our goal is to obtain the nonlinear squeezing at the mode $B$ of the cluster state. To this end, we replace the canonical teleportation protocol with the nonlinear phase cubic measurement \cite{sakaguchi} and, moreover, its ideal theoretical limit of projection onto the cubic state.}
\rev{The POVM element of the projection can be expressed as
\begin{equation}\label{POVM}
    \Pi(m) = C(\chi)\ket{p=m}\bra{p=m}C(\chi)^{\dagger},
\end{equation}
where $\ket{p=m}$ is the eigenstate of the $p$ quadrature with eigenvalue $m$.
In the ideal scenario, the measurement can be understood as a measurement of the nonlinear combination of quadrature operators \eqref{O} that appears in the definition of the nonlinear squeezing \eqref{nlsq}.
The measurement projects on the cubic state \eqref{idfinf}, an eigenstate of the operator $O$ with infinite energy, similar to position eigenstates, which are projected on by homodyne measurement \cite{Gottesman2001,Ghose2007}.} The measurement can be decomposed into the cubic nonlinearity and p-homodyne measurement. These operations can be sequentially applied to one part of the two-mode cluster state as is shown in Fig. \ref{Schemata}b. It is necessary to utilize the measurement result $m$ in a Gaussian feedforward in order to obtain a non-Gaussian state. \rev{If the measurement outcome is discarded, the outcoming state is a mixture of Gaussian states. Here the output is displaced in $p$ over the measurement result with gain $g_p$. In Heisenberg picture, the overall transformation yields
\begin{equation}\label{Poutputgen}
\begin{split}
x_B^{\textrm{out}}=& x_B\\
p_B^{\textrm{out}}=&p_B+g_p p_A+g_p\chi x_A^2.
\end{split}
\end{equation}
Note the appearance of nonlinear terms that originate from the cubic nonlinearity, which action is in more detail described in Appendix \ref{Heisenbergtrasformations}.}

The ideal cubic nonlinearity \eqref{kubik} can be approximated for traveling beams of light by a Gaussian circuit supported by non-Gaussian \rev{auxiliary} state and equipped with \rev{electronic} nonlinear feedforward of the measurement results \cite{Miyata2016,sakaguchi}.\rev{The scheme using this approximation is shown in Fig. \ref{Schemata}d and we call it the nonlinear teleportation as it differs from the canonical scheme only by the nonlinear feedforward.}  The first arm of the bipartite Gaussian state interferes with an \rev{auxiliary} state on a beam splitter. The result of its $x$-homodyne measurement $m_x$ is used in a nonlinear feedforward for a phase shift $\phi = \atan(\sqrt{2}\chi m_x)$ applied to the remaining mode followed by p-homodyne detection. Both measurement results are eventually used to compute the required displacement on the unmeasured mode. Using this circuit, we can asymptotically approach the ideal measurement used in \rev{Fig. \ref{Schemata}b}. The generated transformation of quadratures \rev{is similar} to the one produced by the regular teleportation
\begin{equation}\label{Hnlm}
    \begin{split}
        x_B^{\textrm{out}}&=x_Q+N_x\\
        p_B^{\textrm{out}}&=p'_Q+N'_p.
    \end{split}
\end{equation}
\ver{Whereas the noise term $N_x$ remains the same as in the case of regular teleportation, new nonlinear terms appear in the transformation of the $p$ quadrature}
\begin{equation}
\begin{split}
    N'_p =& p_B+p_A+\frac{\chi}{2}x_A^2\\
    p'_Q =& p_Q-\frac{\chi}{2}x_Q^2.
    \end{split}
\end{equation}
The $p_Q$ quadrature of the \rev{auxiliary} mode Q that initially carries the nonlinear squeezing is nonlinearly transformed and accompanied by \new{non-Gaussian noise terms}. \rev{The transformation already includes the nonlinear feedforward that influences the expression via the parameter $\chi$.}

We further generalize the gains and transmissivity of the beam splitter in the nonlinear teleportation scheme which changes the formulas to
\begin{equation}\label{gentr}
    \begin{split}
        x_B^{\textrm{out}}=&x_B + d_F(tx_Q-rx_A)\\
        p_B^{\textrm{out}}=&p_B + g_F \cos(\theta)(tp_A + rp_Q +\\& \sqrt{2}\chi(tx_A+rx_Q)(rx_A-tx_Q)),
    \end{split}
\end{equation}
with $d_F$ and $g_F$ being electronic gains, and $t$ and $r$ \rev{are} the transmissivity and reflectivity of the beam splitter before the homodyne measurements. By setting $\chi=0$, the equation \eqref{gentr} describes the regular teleportation with generalized gains and beam splitter transmissivity.

To show the positive effect of nonlinear feedforward, let us compare the results obtained by the \rev{canonical} and nonlinear teleportation protocols, \rev{Fig. \ref{Schemata}c and d}. At the same time, we compare the schemes equipped with identical resources of non-Gaussianity \cite{Takagi} and Gaussian squeezing \cite{Braunstein2005b}. The approximations can be classified by the faithful hierarchy of $N$-photon non-Gaussianity \cite{LachmanG19} or stellar rank \cite{chabaud2020}. A superposition \eqref{superpo} with a fixed highest Fock state $N$ has $N$-photon non-Gaussianity. When teleporting a state with some given $N$-photon non-Gaussianity and transferring a state from the same level of the hierarchy by nonlinear teleportation, both schemes \rev{possess} the same $N$-photon non-Gaussian state\rev{,} and any improvement in performance thus originates from the properties of the scheme. Another resource present in the schemes is Gaussian squeezing  \cite{Braunstein2005b}. Throughout our analysis, we limit the available Gaussian squeezing while comparing the performances. At first in preparation of the finitely squeezed state \eqref{idf} \rev{and of the cluster state as well}. We consider a general two-mode Gaussian cluster state, whose parameters are optimized for each situation and each scheme. \rev{Specificaly, we optimize independently $r_1,r_2,\phi,\phi_1,\phi_2,\alpha_1$ and $\alpha_2$ in the preparation of the cluster state as illustrated in Fig. \ref{Schemata}a. Further, we optimize all gains and transitivities of beam splitters in schemes Fig.\ref{Schemata}b,c and d. Eventually, we optimize all free parameters of the  teleported/\rev{auxiliary} states $\chi,r,u,c_0,c_1$ and $c_2$ in \eqref{idf},\eqref{approx2} and \eqref{approx3}} The optimization is done with help of Python library scipy optimize and repeated with randomly chosen initial conditions. Details can be found in Appendix \ref{tmgs}. To remain close to the experimental reality, we consider an \rev{upper bound} on Gaussian squeezing available. The optimum is then sought in a range given by these bounds, e.g. $\pm$9 dB. Overall, by limiting the resources we can perform a genuine comparison of schemes equipped with the same level of $N$-photon non-Gaussianity and limit on maximal Gaussian squeezing.

The parameterization of the two-mode Gaussian state via the covariance matrix can be found in Appendix \ref{Gtms} and is equivalent to two thermal states  transformed by single mode Gaussian operations and a beam splitter as shown in Fig. \ref{Schemata}a. The thermal noise is taken as an input parameter and by its change, we can analyze the resilience of discussed effects. It is described by a parameter $n$ representing the multiples of vacuum fluctuations added to an initially pure vacuum state. \ver{Thus, for a pure state the parameter yields $n=0$ , while $n>0$ represents some amount of thermal noise}  \cite{Mattia2020}.

\subsection{Probabilistic regime}
Schemes described in the previous subsection work in the deterministic regime, accepting all measurement results and processing them in the feedforward. Different measurement results of the homodyne measurements $m_x$ and $m_p$ can generally lead to different quantum states $\hat{\rho}(m_x,m_p)$
\begin{equation}
\hat{\rho}(m_x,m_p) = \frac{\textrm{Tr}_{\textrm{AB}}[\rho_{\textrm{ABQ}}\Pi(m_x,m_p)]}{\textrm{Tr}[\rho_{\textrm{ABQ}}\Pi(m_x,m_p)]}.
\end{equation}
Here $\rho_{\textrm{ABQ}}$ is the density matrix of the cluster state and \rev{the teleported} mode and  $\Pi(m_x,m_p)$ a POVM element describing the measurement. For the ideal nonlinear measurement shown in Fig. \ref{Schemata}b, it is the POVM in \eqref{POVM}. For the regular and nonlinear teleportation, \rev{Fig. \ref{Schemata}c and d,} it yields
\begin{equation}
\begin{split}
    \Pi(m_x,m_p)_c =& U_{BS}(t)^{\dagger}\ket{x=m_x}_Q\bra{x=m_x}\otimes\\& \ket{p=m_p}_A\bra{p=m_p}U_{BS}(t)
\end{split}
\end{equation}
and
\begin{equation}
\begin{split}
    &\Pi(m_x,m_p)_d = U_{BS}(t)^{\dagger}\ket{x=m_x}_Q\bra{x=m_x}\otimes\\& R(\phi(m_x))\ket{p=m_p}_A\bra{p=m_p}R(\phi(m_x))U_{BS}(t),
    \end{split}
\end{equation}
respectively. The operator $U_{BS}(t)$ is a unitary operation of beam splitter with transmissivity $t$ and $R(\phi(m_x))$ a phase shift over angle depending on the measurement outcome $m_x$ as
\begin{equation}
    \re{\phi(m_x) = -\arctan(\chi m_x \sqrt{2})}.
\end{equation}

All states $\rho(m_x,m_p)$ contribute to the outcoming mixture
\begin{equation}\label{mixture}
    \rho_{\textrm{out}}=\int_\sigma P(m_x,m_p) \rho(m_x,m_p) dm_x dm_p,
\end{equation}
where the probability is given by
\begin{equation}
    P(m_x,m_p) = \textrm{Tr}_{\textrm{AB}}[\rho_{\textrm{ABQ}}\Pi(m_x,m_p)].
\end{equation}
\rev{The deterministic scheme is obtained for $\sigma$ being the whole phase space. Generally, the conditionally prepared states $\rho(m_x,m_p)$ are different, and thus, it is advantageous to post-select some of them (i.e., choose $\sigma$ as an appropriate subspace of the phase space) in order to increase the nonlinear squeezing at the output.} \rev{This is, however, accompanied by a reduction in the probability of success.}

\rev{The analysis is performed in the following way. The density matrices of quantum states  $\rho(m_x,m_p)$ are computed for measurement results on a grid of $m_x^i$ and $m_p^i$ with steps $\Delta m_x$ and $\Delta m_p$, such that the trace of their mixture approaches very closely one. Thus, the grid is sufficiently large to capture all states appearing with numerically significant probability. Then, the states $\rho(m_x^i,m_p^i)$ are ordered according to their nonlinear squeezing and aggregated up to some probability $P$. For this mixture of post-selected states, the nonlinear squeezing is evaluated. The corresponding integration domain is $\sigma = \cup_i[(m_x^i,m_x^i + \Delta m_x)\times(m_p^i,m_p^i + \Delta m_p)]$.}

\rev{To conclude with a straightforward example, we consider unity gain linear and nonlinear teleportation via two-mode squeezed vacuum \eqref{esch}.} In fact, the nonlinear feedforward can be more complex, also adjusting the outcoming states $\rho(m_x,m_p)$ by Gaussian operations in such a way, that the nonlinear squeezing is maximized for each of the states. This is equivalent to \rev{the} computation of averaged nonlinear squeezing, where not the conditioned states are aggregated, but their nonlinear squeezing
\begin{equation}\label{aggxi}
    \expval{\xi(z)} = \int_\sigma P(m_x,m_p)\xi(z)_{\rho(m_x,m_p)}dm_xdm_p.
\end{equation}
\rev{In such a way, we present a comparison of the straightforward scheme of unity gain teleportation through a two mode squeezed vacuum with and without the nonlinear feedforward.}

\section{Results}
\subsection{Deterministic scheme}
\begin{figure*}[!htb]
\centering
\includegraphics[width=0.8\textwidth]{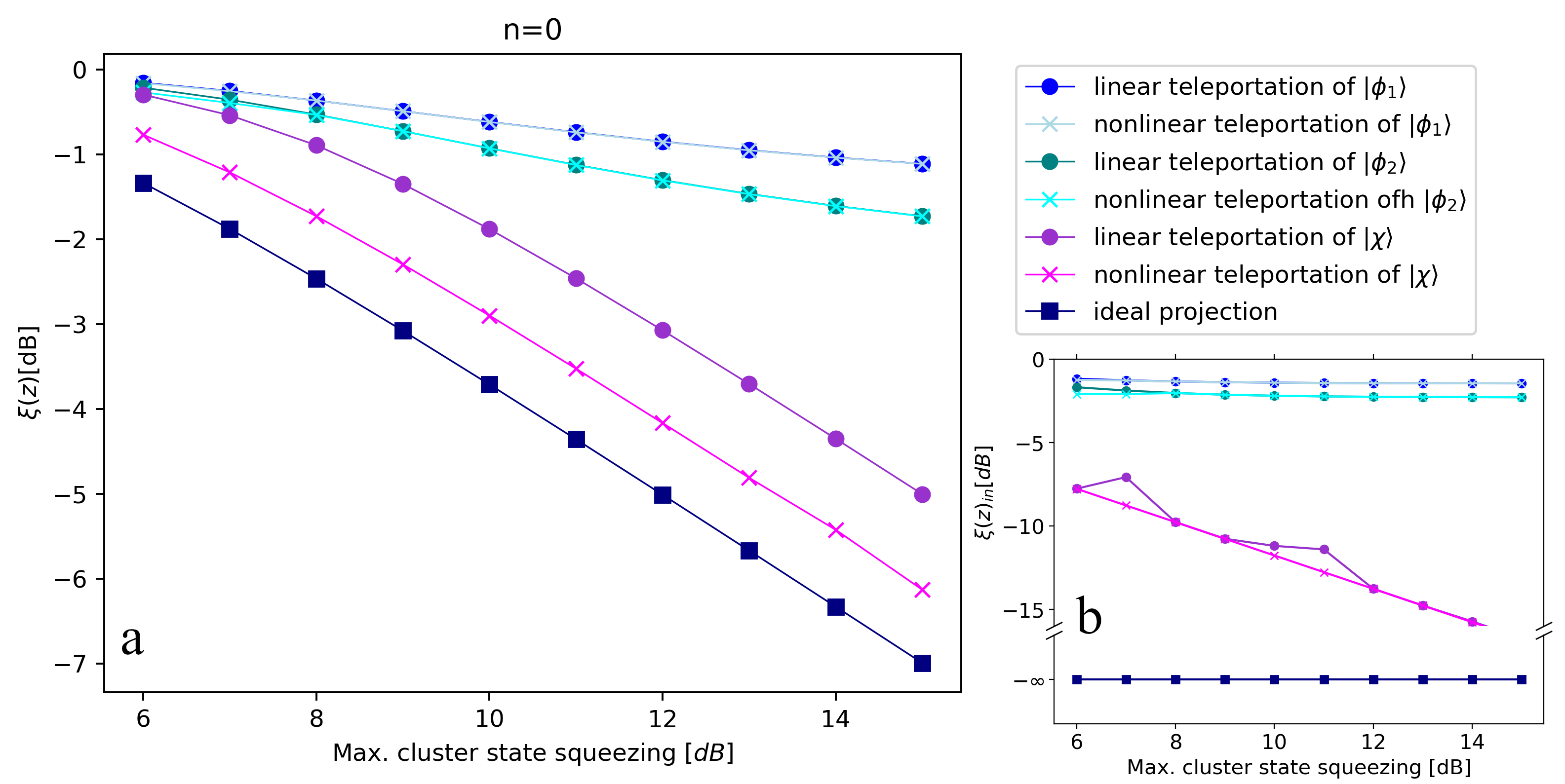}
\caption{\rev{a) Nonlinear squeezing at the output mode, expressed in dB \eqref{xiindB}. The Gaussian squeezing, available for the preparation of the cluster state and the state \eqref{idf}, is limited to a certain value [dB] during optimization. The limit is shown on the horizontal axis.} The two-mode cluster state is pure and optimized for each scheme. Output nonlinear squeezing is shown for \rev{deterministic} regular and nonlinear teleportation \rev{(Fig. \ref{Schemata}c and d)} of the superposition of vacuum state and single photon \eqref{approx2} and the state with 2-photon non-Gaussianity \eqref{approx3}, both with optimized coefficients. \rev{Finally, results are presented for the finitely squeezed cubic state \eqref{idf} serving as teleported/\rev{auxiliary} state and nonlinear cubic phase measurement shown in Fig. \ref{Schemata}b.}  \rev{b) The subplot shows the initial nonlinear squeezing before teleportation/measurement.}}
\label{n0}
\end{figure*}

\rev{Fig. \ref{n0} shows the nonlinear squeezing that can be transferred via an optimized pure two-mode cluster state \rev{($n=0$)} for the individual schemes Fig.\ref{Schemata}b, c, and d. It is plotted as a function of the maximal Gaussian squeezing available for the cluster state preparation and preparation of the state \eqref{idf}, see Appendix \ref{Gtms} for details. Within this limit, the squeezing $r_1$ and $r_2$ forming the cluster state is optimized together with its other parameters $\phi,\phi_1,\phi_2,\alpha_1$, and $\alpha_2$. The cluster state is optimized for each scheme to transfer the maximal amount of the nonlinear squeezing. The teleported or auxiliary states are optimized as well; for example, we optimize the coefficients in superpositions \eqref{approx2} and \eqref{approx3} for best performance, thus obtaining states different \rev{from eigenstates of the restriction} of the $O$ operator on finite Fock subspace.}

Fig. \ref{n0}a can be interpreted pairwise, according to the properties of \rev{the used teleported/auxiliary state and its $1-$ or $2$-photon non-Gaussianity. The state \eqref{approx2}, formed by a superposition of vacuum and the single photon state, has genuine $1-$photon non-Gaussianity. We can compare the performance of the canonical teleportation protocol with linear feedforward Fig.\ref{Schemata}c and the nonlinear teleportation Fig.\ref{Schemata}d when this superposition is used as the teleported state. Thus, the two schemes are initially equipped with a resource of non-Gaussianity of the same level. The teleported state is optimized individually, yielding different $u$ for different schemes (and thus also different initial nonlinear squeezing). Similarly, we can compare the canonical and nonlinear teleportation for the teleported state \eqref{approx3} and, at last, the schemes utilizing a finitely squeezed cubic state \eqref{idf}. In such a pairwise comparison, the initial level of non-Gaussianity is equal, and the only difference is whether the feedforward is linear or nonlinear.}
\rev{Generally, higher squeezing of the cluster state improves the transferred nonlinear squeezing. Fig. \ref{n0}b shows the initial nonlinear squeezing of teleported/auxiliary states. }

\rev{The significant advantage of nonlinear feedforward manifested with the cubic state \eqref{idf} nearly vanishes for the two-component superposition \eqref{approx2}. This shows} the relative strength of linear Gaussian processing of weak non-Gaussian states in \rev{the} deterministic regime. When increasing the resource to two-photon non-Gaussianity  \eqref{approx3} there is a slight difference for a realistic maximal Gaussian squeezing of 6 and 7 dB of the cluster state. A major \ver{improvement by nonlinear feedforward appears} for small Gaussian squeezing when considering finitely nonlinearly squeezed cubic state \eqref{idf}. The difference is even more pronounced for higher \rev{maximal cluster state squeezing. The ideal nonlinear cubic phase measurement, Fig. \ref{Schemata}b, surpasses the performance of all considered schemes. \rev{For comparison, a 10dB cluster state transfers -1.9 dB of nonlinear squeezing via the regular teleportation (Fig. \ref{Schemata}c) and -2.9 dB via the nonlinear teleportation (Fig. \ref{Schemata}d) of the finitely squeezed cubic state and -3.7 dB with the ideal projection (Fig. \ref{Schemata}b).} Interestingly, the slope with which the nonlinear squeezing improves with increasing cluster state squeezing is comparable for all three schemes, when the cubic state is involved, or projected on. Altogether, when increasing to $2$-photon non-Gaussianity of the supporting state \eqref{approx3}, the nonlinear teleportation with nonlinear feedforward starts manifesting an advantage over teleportation, which lies in higher nonlinear squeezing at the output.}

\begin{figure*}[!htb]
\centering
\includegraphics[width=0.8\textwidth]{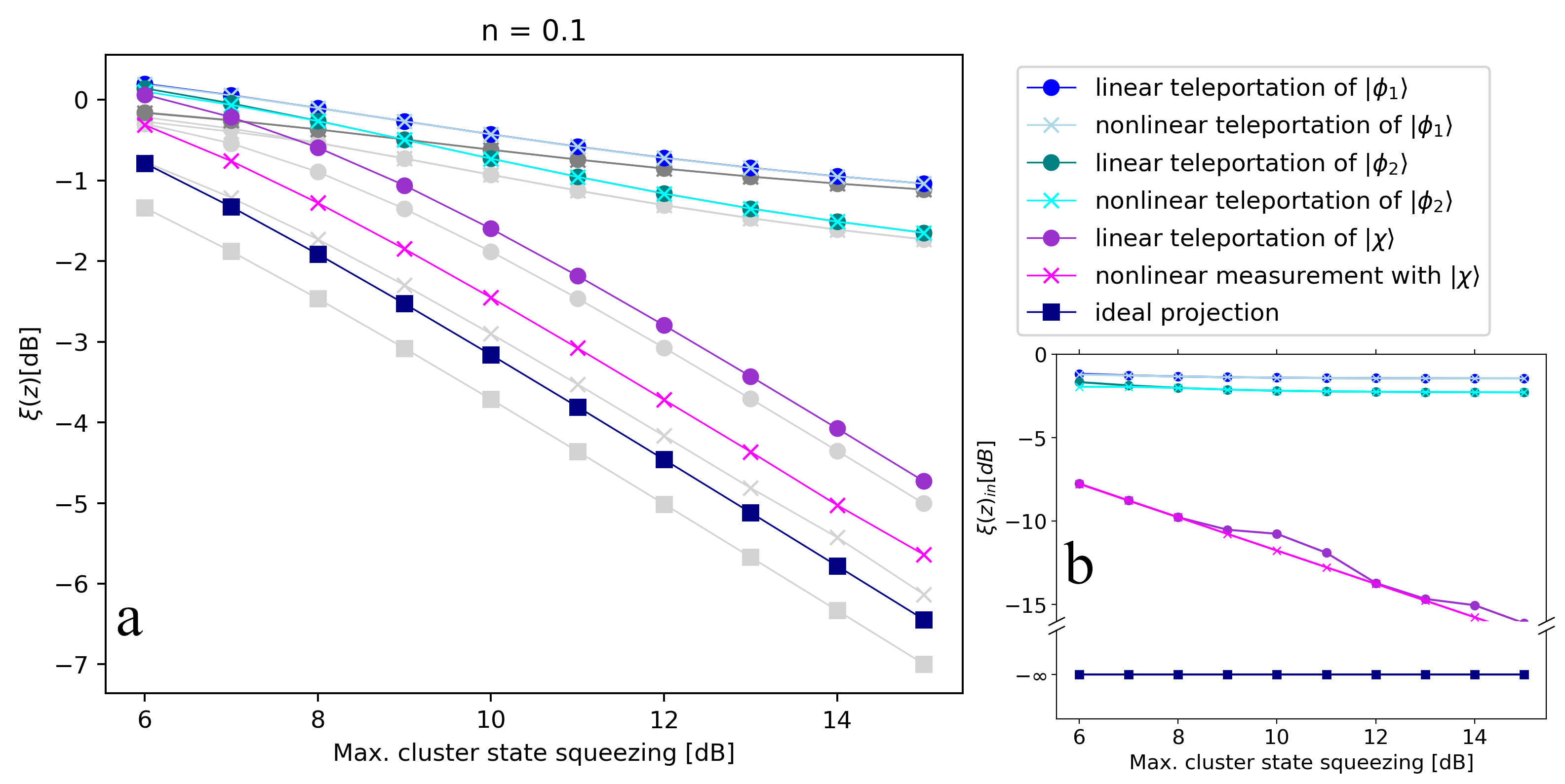}
\caption{\rev{a) Nonlinear squeezing at the output mode, expressed in dB \eqref{xiindB}. The Gaussian squeezing, available for the preparation of the cluster state and the state \eqref{idf}, is limited to a certain value [dB] during optimization. The limit is shown on the horizontal axis.} The two-mode cluster state carries $n=0.1$ additional thermal noise. \rev{To better see the comparison with the case of a pure cluster state (Fig. \ref{n0}a), the light gray shadow shows results for $n=0$.}\rev{b) The subplot shows the initial nonlinear squeezing before teleportation/measurement.}}
\label{n1}
\end{figure*}

Fig. \ref{n1}a shows nonlinear squeezing transferred via the schemes in Fig. \ref{Schemata}b,c, and d when the cluster state is initially prepared not from a pure vacuum state but from thermal states that carry $n = 0.1$ additional fluctuations over the vacuum. The following optimization of the cluster state and the schemes is as for the previous plot within bounded Gaussian squeezing. The performance of all schemes is negatively affected; the nonlinear squeezing transferred with the pure cluster state is depicted with the light shadow. The relative comparison of the schemes remains unchanged. \rev{Fig. \ref{n1}b shows the initial nonlinear squeezing. } 

\subsection{Probabilistic regime}
\rev{In the previous section, we showed the potential of the nonlinear feedforward. In the deterministic regime, the advantage is more pronounced for the experimentally challenging cubic state. Here, we will turn our attention to the probabilistic regime, seeking a case that enables showing the advantage of the nonlinear feedforward even with currently accessible states.} Let us start with \rev{a} more detailed discussion of some cases presented in Fig. \ref{n0}\rev{a}. \rev{To fully understand the role of the nonlinear feedforward, we will look at teleportation via 6 dB, 7 dB, and 8 dB cluster states in a probabilistic regime at all the measurement outcomes individually.} We choose the \rev{two-component} superposition \eqref{approx2} for this analysis as it is close to experimental reality, and the parameters of optical schemes and input states are taken from the optimized deterministic regime. \rev{Thus} they are the same as for which the results shown in Fig. \ref{n0} were obtained.

\begin{figure*}[!htb]
\centering
\includegraphics[width=0.8\textwidth]{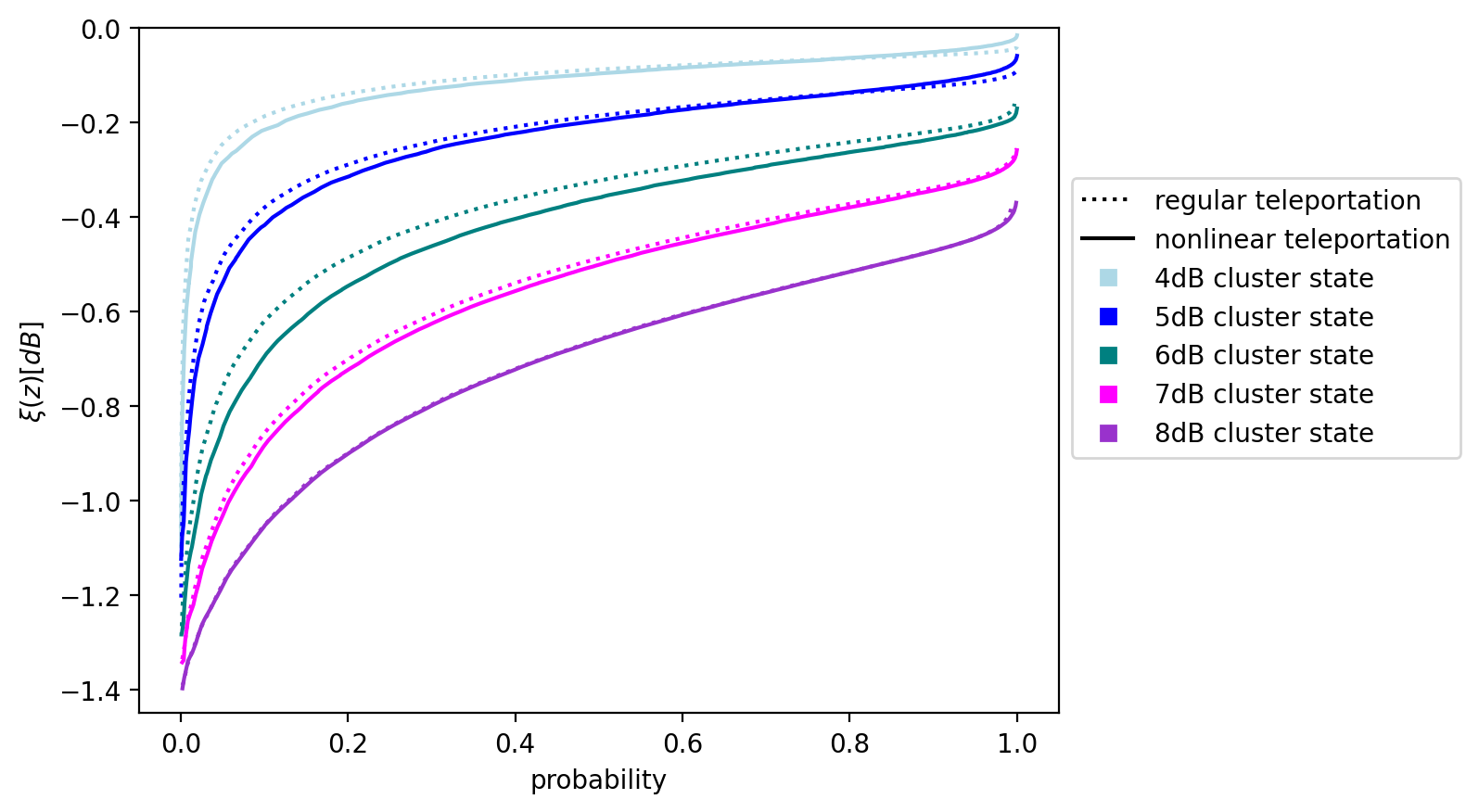}
\caption{\ver{Nonlinear squeezing teleported by linear (dotted line) and nonlinear (solid line) teleportation. The input state and parameters of the optical scheme are optimized for the deterministic scenario. Quantum states depending on the measurement results with best nonlinear squeezing are aggregated up to a given probability. }}
\label{aggdet}
\end{figure*}

\rev{As more measurement results are accepted and therefore, the output mixture includes} more states, the overall probability of success or obtaining such a state increases, whereas the nonlinear squeezing diminishes, as can be seen in Fig. \ref{aggdet}. This happens as a result of accepting less nonlinearly squeezed states to the mixture. As we increase the squeezing of \rev{the} cluster state, the advantage brought by the nonlinear feedforward vanishes. \rev{To see the behavior as we decrease the cluster state squeezing, we added a case of 4dB and 5dB of squeezing in the cluster state; however, from discussed cases, teleportation through the 6dB cluster state manifests the largest difference provided by the nonlinear feedforward. 

Measurement results around the origin of the phase space alter the teleported state only slightly, which diminishes the difference between the linear and nonlinear feedforward in schemes Fig. \ref{Schemata}c and d. Increasing the probability means including larger measurement results. At a certain point, a discrepancy appears between the nonlinear feedforward designed for the ideal cubic state and the auxiliary state, which is the two-component superposition. This leads to similar performance of the canonical teleportation and nonlinear measurement at higher probabilities. The difference is best seen for the success probability $0.2$ in Fig. \ref{aggdet}. }

\begin{figure*}[!htb]
\centering
\includegraphics[width=0.9\textwidth]{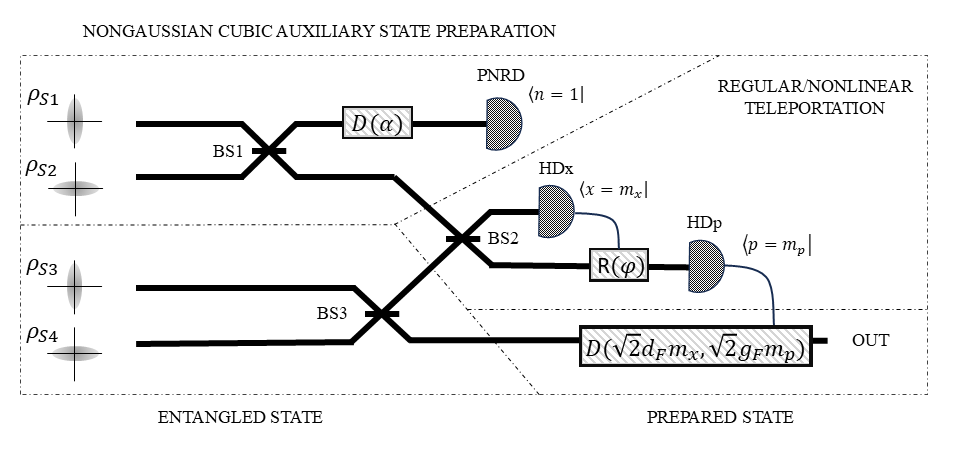}
\caption{Scheme for conditioned regular/nonlinear teleportation \rev{with the two-mode squeezed vacuum} including the preparation of the \rev{teleported} state. \rev{The outcoming state is accepted or discarded based on the measurement results $m_x$ and $m_p$.}}
\label{esch}
\end{figure*}

We conclude our analysis by comparing \rev{the} performance of unity gain teleportation with and without the nonlinear feedforward\rev{, shown in Fig \ref{esch}}. As the teleported state is considered the experimentally realistic \rev{two-component} superposition \eqref{approx2}. Its specific form given by $u = 0.79$ is chosen such that the state is \rev{an} eigenstate of the operator $O$ in Eq. \eqref{O} restricted in Fock space to the subspace of dimension two. This approach ensures maximal nonlinear squeezing that can be obtained with \rev{the superposition of vacuum and single-photon state.} The nonlinear squeezing is evaluated at this \rev{state's} native cubicity \rev{$z_n$ \eqref{nativez}}. \rev{Thus, a state with maximal nonlinear squeezing on a Fock subspace of dimension two \eqref{approx2} is teleported, and afterward, evaluation is done at its native cubicity before the teleportation.} This setting leaves us with a single degree of freedom, $\chi$ in the nonlinear feedforward, which is optimized and yields $\chi=-0.219$ in the case of 6dB cluster state and  $\chi = -0.130$ for 8dB cluster state.

\begin{figure*}[!htb]
\centering
\includegraphics[width=0.8\textwidth]{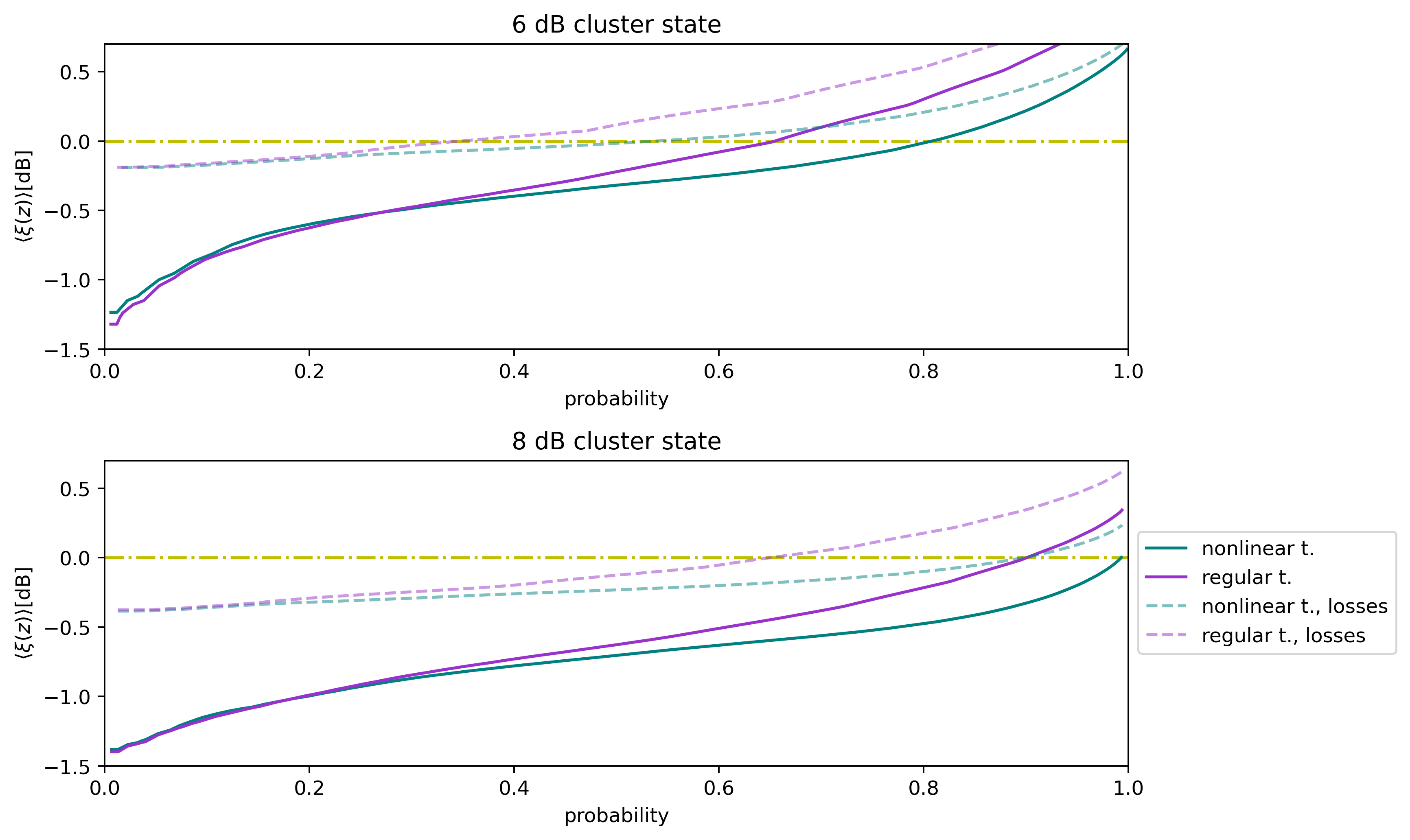}
\caption{Nonlinear squeezing \eqref{aggxi} produced by unity gain regular/nonlinear teleportation of an eigenstate of the operator in Eq. $O$ \eqref{O} conditioned on measurement results of the homodyne measurement with best results aggregated up to some probability and evaluated at native cubicity of the state before teleportation. The nonlinear teleportation and regular teleportation \rev{were} simulated with two mode squeezed vacuum of 6 dB and 8 dB with (dashed lines) and without losses on the
teleported state. Here it is assumed $25\%$ losses \cite{Konno2021}. In all cases the nonlinear squeezing \rev{vanishes} first (the plot line
crosses the dashed line of boundary 1) for regular teleportation.}
\label{agg}
\end{figure*}

Nonlinear squeezing shown in Fig. \ref{agg} is visibly enhanced by nonlinear feedforward and enables its detection even for probabilities for which the regular teleportation fails to transfer any. We also introduced $25\%$ losses to the \rev{auxiliary} state, \rev{which} was present in its recent experimental preparation. The results are shown in dashed lines and again show \rev{the} advantage of the nonlinear teleportation with nonlinear feedforward. The optimal strength of the nonlinear feedforward is higher than for the ideal case, yielding $\chi = -0.250$ for 6dB cluster state and  $\chi = -0.151$ for 8dB cluster state. A similar effect was recognized in \rev{the} decoherence of the nonlinear squeezing \cite{Kala22}, where the native cubicity of the state increases with higher losses. In the presence of losses, the performances of teleportations through cluster states with 6 and 8 dB of Gaussian squeezing do not converge even for postselection on the best cases with vanishing probability.

\section{Conclusion}
The great remaining challenge of photonic continuous variable quantum computing remains in accompanying the rich multimode structure of Gaussian cluster states by non-Gaussian elements. As the basic mechanism of such cluster-state quantum computing serves the canonical teleportation protocol. Its performance, when non-Gaussianity is present, can be improved by adapting the measurement results processing.

\rev{Instead of a teleportation of the typical Fock states \cite{Ogawa2016} or cat states \cite{Ourjoumtsev2006}, here, we consider cubic nonlinear squeezing, a notion of purely continuous-variable non-Gaussianity, that has a straightforward interpretation as a resource for nonlinear quantum operations \cite{Lloyd1999,Budinger2024}.} We showed an improvement in the teleported nonlinear squeezing by involving currently experimentally tested nonlinear feedforward \cite{sakaguchi}. In a deterministic regime, the positive effect can be seen on the theoretical level of teleportation of the ideal cubic state. However in a probabilistic regime, when best results are postselected up to some probability, the improvement is observable even with current experimental resources in comparison to unity gain teleportation.
Processing the measurement data during teleportation via classical nonlinear feedforward decreases the added
amount of noise which can be seen by improved nonlinear squeezing at the output. The cubic nonlinear squeezing is a resource for measurement-based quantum computation \cite{Konno2021,sakaguchi} and \rev{improving how much of it is transferred via teleportation increases the potential to chain and iterate computational steps as required by cluster computing.  }\rev{The cubic nonlinearity itself is not currently experimentally available in traveling light. However,} a superposition of Fock states with nonlinear squeezing has already been prepared \cite{Konno2021} \new{as an alternative resource substituting the nonlinearity and} \rev{a design based on the Kerr nonlinearity is also proposed for near future tests \cite{Yana2020}.} 

\rev{Secondly, we show that in a probabilistic regime, the nonlinear teleportation equipped with nonlinear feedforward and currently experimentally accessible state outperforms the unity gain linear teleportation through 6dB cluster state as is shown in Fig. \ref{agg}. The advantage is thus, in principle, observable even with the currently available nonlinearly squeezed state and appears by adjusting the classical part of the scheme consisting of nonlinear feedforward combined with post-selection.} The enhancement remains observable even in the presence of losses that accompanied the experimental preparation \cite{Konno2021}.
\ver{Our work is one of the first steps required to extend the presently Gaussian theory of cluster states into its final destination - \rev{the} non-Gaussian domain.}

\section*{Acknowledgement}
VK and PM acknowledge grant 22-08772S and RF grant 21-13491X of Czech Science Foundation. VK acknowledges the Quantera project CLUSSTAR (8C24003) of MEYS, Czech Republic. Project CLUSSTAR has received funding from the European Union’s Horizon 2020 Research and Innovation Programme under Grant Agreements No. 731473 and No. 101017733 (QuantERA). PM acknowledges a grant from the Programme Johannes Amos Comenius under the Ministry of Education, Youth and Sports of the Czech Republic reg. no. CZ.02.01.01/00/22\textunderscore 008/0004649. VK and PM  also acknowledge Horizon Europe Research and Innovation Actions under Grant Agreement no. 101080173 (CLUSTEC). M.W. acknowledges funding from the Plan France 2030 through the project ANR-23-PETQ-0013. VK acknowledges project IGA-PrF-2025-008.

\section*{Data availability}
The data that support the findings of this article are openly available at [10.5281/zenodo.18402580].

\appendix

\section{Model in Heisenberg picture}\label{Heisenbergtrasformations}
In order to evaluate the nonlinear squeezing at the output of presented schemes we compute the transformation \new{they generate} in the Heisenberg picture.
At first we describe the action of optical elements present in the schemes as quadrature transformations. The beam splitter generates a two-mode transformation on quadratures $x_1,p_1,x_2$ and $p_2$ of the form
\begin{equation}\label{BS}
    \begin{split}
        x_1\rightarrow& tx_1+rx_2\\
        p_1 \rightarrow& t p_1+rp_2\\
        x_2 \rightarrow& tx_2-rx_1\\
        p_2 \rightarrow& tp_2-rp_1,
    \end{split}
\end{equation}
where $t$ and $r$ are transmissivity and reflectivity, respectively. Equation $t^2+r^2=1$ holds.

Cubic nonlinearity acts on quadratures $x$ and $p$ as
\begin{equation}
    \begin{split}
        x\rightarrow& x\\
        p \rightarrow& p+\chi x^2,
    \end{split}
\end{equation}
where we can see the appearance of the quadratic non-Gaussian term.
Measurement of a quadrature $q$ with outcome $m$ can be expressed as
\begin{equation}
    q=m,
\end{equation}
provided, that the measurement outcome is used for feedforward to obtain independence on $m$.\\

Lets consider a general two-mode Gaussian state defined by quadratures $x_A,p_A,x_B$ and $p_B$, where the $A$ mode undergoes an ideal  cubic nonlinearity followed by a homodyne measurement of $p$ with measurement result proportional to
\begin{equation}
    i_p = p_A + \chi x_A^2.
\end{equation}
The measurement result is further used for $p$ displacement of the mode $B$ with gain $g_p$, yielding
\begin{equation}\label{Poutputgengen}
\begin{split}
x_B^{\textrm{out}}=& x_B\\
p_B^{\textrm{out}}=&p_B+g_p p_A+g_p\chi x_A^2+zx_B^2.
\end{split}
\end{equation}

The nonlinear variance in \eqref{nlsq} can be in this case expressed  in terms of the quadratures of the two Gaussian modes $A$ and $B$ yields
\begin{equation}
\begin{split}
&\myvar(p_B^{\textrm{out}}+z x_B^{\textrm{out}2})=\\
&\myvar(p_B)+2g_p \mycov(p_A,p_B)+g_p^2\myvar(p_A)\\
&+2\chi g_p \mycov(p_B,x_A^2) + 2z\mycov(p_B,x_B^2)\\
&+2g_p^2\chi \mycov(p_A,x_A^2)+2g_p z \mycov(p_A,x_B^2)\\
&+g_p^2 \chi^2 \myvar(x_A^2)+2g_p \chi z \mycov(x_A^2,x_B^2)+z^2\myvar(x_B^2).
\end{split}
\end{equation}

The schemes of measurement-induced nonlinearity and teleportation can be described by one model as they differ only by feedforwarded rotation parameterized by $\chi$ which can be switched of in the case of teleportation by setting $\chi=0$.
The measurement results of $x$-homodyne and p-homodyne in c) and d) of Fig. \ref{Schemata} yields
\begin{equation}
\begin{split}
    m_x =& t x_Q - r_xA\\
    m_p =& \cos(\theta)(tp_A + rp_Q) -\sin(\theta)(tx_A + rx_Q),
\end{split}
\end{equation}
where quadratures with subscript $Q$ are of the \rev{auxiliary} mode and with $A$ of the first mode of the cluster state; $\theta$ depends on the measurement outcome of $x$-homodyne as
\begin{equation}
    \theta = \sqrt{2}m_x\chi.
\end{equation}
The measurement results are further feedforwarded to displacement on the output mode B with gains $g_F$ and $d_F$
\begin{equation}
\begin{split}
    x_B^{\textrm{out}} = d_F m_x\\
    p_B^{\textrm{out}} + g_F m_p,\\
\end{split}
\end{equation}
where the superscript out denotes quadratures the the output of the scheme. We can now write the whole transformation
\begin{equation}
    \begin{split}
        x_B^{\textrm{out}}=&x_B + d_F(tx_Q-rx_A)\\
        p_B^{\textrm{out}}=&p_B + g_F \cos(\theta)(tp_A + rp_Q +\\& \sqrt{2}\chi(tx_A+rx_Q)(rx_A-tx_Q))
    \end{split}
\end{equation}
Using this transformation and \new{considering the second moments of modes A and B to be given by covariance matrices in \eqref{covariancematg}}, the variance of the operator $p_B^{\textrm{out}} + zx_B^{\textrm{out}2}$ can be expressed as follows.
With shorthand notation of coefficients
\begin{equation}
    \begin{split}
        A =& z\\
        B =& 2d_Ftz\\
        C =& -2d_Frz\\
        D =& -2d_F^2rtz - \alpha\chi g_F t^2 + \alpha \chi g_F r^2\\
        E =& d_F^2r^2z + \alpha \chi g_Frt\\
        F =& d_F^2t^2z - \alpha \chi g_F r t
    \end{split}
\end{equation}
we can write
\begin{equation}
    \begin{split}
        &\myvar(p_B^{\textrm{out}} + z x_B^{\textrm{out}2})=\\
        &\myvar(p_B) + (\alpha g_F t)^2 \myvar(p_A) + (\alpha g_F r)^2\myvar(p_Q)+\\&2\alpha g_F t \mycov(p_A,p_B)
        2\{ A \mycov(p_B,x_B^2) + B\mycov(p_B,x_Qx_B) +\\&C\mycov(p_B,x_Ax_B)
        +D\mycov(p_B,x_Ax_Q) + E\mycov(p_B,x_A^2)+\\
        &\alpha g t (A\mycov(p_A,x_B^2) +B\mycov(p_A,x_Qx_B)+\\&C\mycov(p_A,x_Ax_B) + 
        D\mycov(p_A,x_Ax_Q)
        +E\mycov(p_A,x_A^2))\\
        &+\alpha g r (B\mycov(p_Q,x_Qx_B) + D\mycov(x_Ax_Q,p_Q) + \\&F\mycov(p_Q,x_Q^2)\}
        A^2\myvar(x_B^2)+ 2AB \mycov(x_B^2,x_Qx_B) + \\&2AC \mycov(x_B^2,x_Ax_B) + 
        2AD\mycov(x_B^2,x_Ax_Q)
        +\\&2AE\mycov(x_A^2,x_B^2)
        + B^2\myvar(x_Qx_B)\\&+2BC\mycov(x_Qx_B,x_Ax_B)(-2drz)
        + \\&2BD\mycov(x_Qx_B,x_Ax_Q)
        +2BE\mycov(x_Qx_B,x_A^2) + \\&2BF\mycov(x_Qx_B,x_Q^2) + F^2\myvar(x_Q^2)
        2FC\mycov(x_Q^2,x_Ax_B)+ \\&2FD\mycov(x_Q^2,x_Ax_Q) + 
        2FE\mycov(x_Q^2,x_A^2)
        +\\&C^2\myvar(x_Ax_B) +2CD\mycov(x_Ax_B,x_Ax_Q)
        +\\&2CE\mycov(x_Ax_B,x_A^2) +D^2\myvar(x_Ax_Q)
        +\\&2DE\mycov(x_Ax_Q,x_A^2)
        +E^2\myvar(x_A^2).
    \end{split}
\end{equation}
The Gaussianity of the cluster state can be used for expressing the whole expression in terms appearing in covariance matrix of the modes A and B. For the more complex moments involving more than two quadratures, we use the generalized Stein's lemma for multivariate Gaussian distributions of a random variable $\Vec{X}$ with mean values $\Vec{\mu}$ \cite{Liu1994}
\begin{equation}
    E(g(X)X_i) = \mu_iE(g(X))+\sum_jC_{ij}E(\partial_jg(X)).
\end{equation}
The Stein's lemma cannot be used generally for computation of quantum moments. It is necessary, that those moments are Weyl symmetrical.

\begin{equation}
    \begin{split}
        &\mycov(p_B,x_B^2) = 2\expval{x_B}\mycov(x_P,p_B)\\&\mycov(p_B,x_Qx_B) = 0 \textrm{ due to} \expval{x_Q}=0\\
        &\mycov(p_B,x_Ax_Q) = \expval{x_Q}\mycov(x_A,p_B)=0\\
        &\mycov(p_B,x_A^2) = 2\expval{x_A}\mycov(x_A,p_B)\\
        &\mycov(p_A,x_B^2) = 2\expval{x_B}\mycov(x_B,p_A)\\
        &\mycov(p_A,x_Qx_B) = 0\\
        &\mycov(p_A,x_Ax_Q) = 0\\
        &\mycov(p_A,x_A^2) = 2\expval{x_A}\mycov(x_A,p_A)\\
        &\mycov(p_Q,x_Qx_B) = \expval{x_B}\mycov(x_Q,p_Q)\\
        &\mycov(p_Q,x_Ax_Q) = \expval{x_A}\mycov(x_Q,p_Q)\\
        &\mycov(p_B,x_Ax_B) = \expval{x_B}\mycov(x_A,p_B)+ \expval{x_A}\mycov(x_B,p_B)\\
        &\mycov(p_A,x_Ax_B) = \expval{x_A}\mycov(x_B,p_A)+\expval{x_B}\mycov(x_A,p_A)\\
        &\mycov(x_B^2,x_Qx_B) = 0\\
        &\mycov(x_B^2,x_Ax_B) = \expval{x_A}\expval{x_B}^3 + 3\mycov(x_A,x_B)\expval{x_B^2}+ \\&3 \myvar(x_B)\expval{x_A}\expval{x_B}+ 3\myvar(x_B)\mycov(x_A,x_B)-(\mycov(x_A,x_B)+ \\&\expval{x_A}\expval{x_B})(\myvar(x_B)+\expval{x_B}^2)\\
        &=2\expval{x_B}^2\mycov(x_A,x_B) + 2\expval{x_A}\expval{x_B}\myvar(x_B) +\\& 2\myvar(x_B)\mycov(x_A,x_B)\\
        &\mycov(x_B^2,x_Ax_Q) = 0\\
        &\myvar(x_Qx_B) = \expval{x_Q^2}(\myvar(x_B)+ \expval{x_B}^2)\\
        &\mycov(x_A^2,x_B^2) = 4\mycov(x_A,x_B)\expval{x_A}\expval{x_B} + 2\mycov(x_A,x_B)^2\\
        &\mycov(x_Qx_B,x_Ax_B) = \expval{x_Q}\expval{x_B}\mycov(x_A,x_B) + \\&\expval{x_Q}\expval{x_A}\mycov(x_B,x_B)\\
        &\mycov(x_Qx_B,x_A^2) = 0\\
        &\mycov(x_Qx_B,x_Q^2) = \expval{x_B}\mycov(x_Q,x_Q^2)\\
        &\mycov(x_Q^2,x_Ax_B) = 0\\
        &\mycov(x_Q^2,x_Ax_Q) = \expval{x_A}\mycov(x_Q,x_Q^2)\\
        &\mycov(x_Q^2,x_A^2) = 0\\
        &\myvar(x_Ax_B) = \myvar(x_A)\expval{x_B}^2 + 2\mycov(x_A,x_B)\expval{x_A}\expval{x_B}+\\&\myvar(x_B)\expval{x_A}^2 + \mycov(x_A,x_B)^2+\myvar(x_A)\myvar(x_B)\\
        &\mycov(x_Ax_B,x_Ax_Q) = 0\\
        &\mycov(x_Ax_B,x_A^2) = \expval{x_B}\expval{x_A}^3 + 3\mycov(x_A,x_B)\expval{x_A}^2 + \\&3\myvar(x_A)\expval{x_A}\expval{x_B} + 3 \myvar(x_A)\mycov(x_A,x_B)-(\mycov(x_A,x_B)+\\&\expval{x_A}\expval{x_B})(\myvar(x_A)+ \expval{x_A}^2)\\
        &=2(\myvar(x_A)\mycov(x_A,x_B) + \mycov(x_A,x_B)\expval{x_A}^2 +\\& \myvar(x_A)\expval{x_A}\expval{x_B})\\
        &\myvar(x_Ax_Q) = \expval{x_Q^2}(\myvar(x_A)+ \expval{x_A}^2)\\
        &\mycov(x_Ax_Q,x_A^2) = 0\\
        &\myvar(x_A^2) = 4\myvar(x_A)\expval{x_A}^2 + 2\myvar(x_A)^2\\
        &\myvar(x_B^2) = 4\myvar(x_B)\expval{x_B}^2 + 2\myvar(x_B)^2
    \end{split}
\end{equation}
Additionally, for the finitely nonlinearly squeezed state \eqref{idf} and its approximation \eqref{approx2} holds 
\begin{equation}\label{assumption}
    \begin{split}
        \textrm{cov}(x_Q,p_Q)=0&\\
        \expval{x_Q}=0&\\
        \textrm{cov}(x_Q,x_Q^2)=0&\\
        \expval{x_Q^2}=\frac{1}{2g}
    \end{split}
\end{equation}
\begin{equation}\label{assumptionstate}
    \expval{x_Q^2}=\frac{3}{2} - u^2
\end{equation}
\begin{equation}\label{assumptionanc}
    \begin{split}
        \textrm{cov}(x_Q,p'_Q)=0&\\
        \expval{x_Q}=0&\\
        \textrm{cov}(x_Q,x_Q^2)=0&\\
        \expval{x_Q^2}=\frac{3}{2}-u^2
    \end{split}
\end{equation}

\begin{equation}
    \begin{split}
    &\myvar(p_Q) = 2u^4-3u^2+\frac{3}{2}\\
    &\myvar(x_Q^2) = \frac{3}{2} - u^4\\
    &\mycov(x_Q^2,p_Q) = \sqrt{2(1-u^2)}u(u^2-1).
    \end{split}
\end{equation}

\section{Parametrization of the general two-mode cluster state and its optimization}\label{Gtms}

For the two-mode cluster state we considered a general  Gaussian state which is prepared from a thermal state $\rho_T$ as
\begin{equation}\label{gentms}
    \begin{split}    &R_1(\phi_1)R_2(\phi_2)U_{\textrm{BS}}R_1(\phi)S_1(r_1)S_2(r_2)\rho_T\otimes\\& \rho_T S_2(r_2)^{\dagger}S_1(r_1)^{\dagger}R_1(\phi)^{\dagger}U_{\textrm{BS}}^{\dagger}R_2(\phi_2)^{\dagger}R_1(\phi_1)^{\dagger},
    \end{split}
\end{equation}
where $R_i(\phi),S_i(r)$ are rotation and Gaussian squeezing acting on the $i-$th mode and $U_{\textrm{BS}}$ is a unitary transformation of beam splitter.
Initial thermal state has covariance matrix
\begin{equation}
\gamma_{\textrm{in}}= \begin{pmatrix}
\frac{1}{2}(1+n)&0&0&0\\
0&\frac{1}{2}(1+n)&0&0\\
0&0&\frac{1}{2}(1+n)&0\\
0&0&0&\frac{1}{2}(1+n).
\end{pmatrix}
\end{equation}
Being a pure state for $n=0$ and having $n$ times added vacuum noise if thermal.
The symplectic transformation representing squeezing yields
\begin{equation}\label{cmat1}
S= \begin{pmatrix}
\frac{1}{g_1}&0&0&0\\
0&g_1&0&0\\
0&0&g_2&0\\
0&0&0&\frac{1}{g_2}
\end{pmatrix},
\end{equation}
where $g=\exp(r/2)$.
Symplectic transformation of beam splitter
\begin{equation}\label{cmat2}
S_{BS}= \begin{pmatrix}
t&0&r&0\\
0&t&0&r\\
-r&0&t&0\\
0&-r&0&t
\end{pmatrix},
\end{equation}
where $t$ is transmissivity and $t^2+r^2=1$ holds.
Rotation of the first mode is described as
\begin{equation}\label{cmat3}
R_1= \begin{pmatrix}
\cos(\phi)&\sin(\phi)&0&0\\
-\sin(\phi)&\cos(\phi)&0&0\\
0&0&1&0\\
0&0&0&1
\end{pmatrix}.
\end{equation}
And finally, the rotation of both modes yields
\begin{equation}\label{cmat4}
R_{12}= \begin{pmatrix}
\cos(\phi_1)&\sin(\phi_1)&0&0\\
-\sin(\phi_1)&\cos(\phi_1)&0&0\\
0&0&\cos(\phi_2)&\sin(\phi_2)\\
0&0&-\sin(\phi_2)&\cos(\phi_2)
\end{pmatrix}.
\end{equation}
The overall covariance matrix of the two-mode state is then computed as
\begin{equation}\label{covariancematg}
    \gamma = R_{12}S_{\textrm{BS}}R_1S\gamma_{\textrm{in}}S^T R_1^T S_{\textrm{BS}}^T R_{12}^T
\end{equation}

\section{Optimization of the cluster state}\label{tmgs}
We performed numerical optimization with scipy.optimize minimize modul with 400-460 randomly chosen initial states with covariance matrix \eqref{covariancematg}. The initial conditions were chosen within a range given by the upper bound on the maximal Gaussian squeezing,  an interval $\pm 10$ for mean values of $x$ quadratures, cubicities $\chi$ and $z$ and gains in an interval $\pm 2$ and arbitrary rotations.

\bibliography{references}

\end{document}